# Self-organizing GeV nano-Coulomb collimated proton beam from laser foil interaction at $7 \times 10^{21}$ W/cm$^2$

X.Q. Yan[1,2,4*], H.C. Wu[1], Z.M. Sheng[3], J.E. Chen[2], J. Meyer-ter-Vehn[1]

[1] Max-Planck-Institut fuer Quantenoptik, Hans-Kopfermann-Straße 1, D-85748 Garching, Germany

[2] State Key Laboratory of Nuclear Physics and Technology, Peking University, Beijing 100871, China

[3] Beijing National Laboratory for Condensed Matter Physics, Institute of Physics, CAS, Beijing 100190, China, and Department of Physics, Shanghai Jiao Tong University, Shanghai 200240, China.

[4] Center for Applied Physics and Technology, Peking University, Beijing 100871, China

Abstract:

We report on a self-organizing, quasi-stable regime of laser proton acceleration, producing 1 GeV nano-Coulomb proton bunches from laser foil interaction at an intensity of $7 \times 10^{21}$ W/cm$^2$. The results are obtained from 2D PIC simulations, using circular polarized laser pulse with Gaussian transverse profile, normally incident on a planar, 500 nm thick hydrogen foil. While foil plasma driven in the wings of the driving pulse is dispersed, a stable central clump with 1 - 2 λ diameter is forming on the axis. The stabilisation is related to laser light having passed the transparent parts of the foil in the wing region and encompassing the still opaque central clump. This feature is observed consistently in 2D and 3D simulations. It depends on a laser pulse shape with high contrast ratio.

[*] Email: xyan@mpq.mpg.de

With the development of chirped pulse amplification technique [1], ultra-intense short laser pulses with peak intensity as high as $I > 10^{21}$ W/cm$^2$ and contrast ratios in excess of $10^{10}$ are now available, allowing for studies of laser interaction with ultrathin targets [2]. Energetic ions can be produced by means of intense laser light interacting with thin foils. These ion beams are attracting much attention due to a wide range of potential applications covering radiograph transient processes [3], ion beam tumor therapy [4], and fast ignition of fusion cores [5]. As a rule, these applications require ion beams with low energy spread and high collimation.

Usually a linear polarized (LP) laser pulse is used which generates hot electrons due to J×B heating [6]. Ion acceleration mechanisms [7–11] are electrostatic shock acceleration at the front and target normal sheath acceleration (TNSA) at the rear side. However, the proton pulses obtained in this way are far from monochromatic. Often the energy spread is 100%, and ions with the highest energy represent only a small fraction of the total flux [12]. Certain techniques can be used to decrease the energy spread [13-15], and the best results to date have yielded an energy spread of about 20% FWHM at relatively low energy (few MeV).

In the present paper, we consider circular polarized (CP) light. As pointed out recently in a number of papers [16-20], CP laser pulses can accelerate ions very efficiently and produce sharply peaked spectra. When normally incident on plane foils, the light pressure is quasi-stationary, following only the time dependence of the pulse envelope. Electrons are then smoothly pushed into the high-density material without strong heating and ions are taken along by means of the charge separation field. This is in contrast to linear polarization which triggers fast longitudinal electron oscillations and excessive heating.

For appropriate parameters, CP pulses may accelerate foils as a whole with most of the transferred energy carried by ions. The basic dynamics are well described by a one-dimensional (1D) piston model [18, 20]. Acceleration terminates due to multi-dimensional effects such as transverse expansion of the accelerated ion bunch and transverse instabilities. In particular, instabilities grow in the wings of the indented foil, where light is obliquely incident and strong electron heating sets in. Eventually, this part of the foil is diluted and becomes transparent to the driving laser light. The central new observation in the present paper is that this process of foil dispersion may stop before reaching the centre of the focal spot and that a relatively stable ion clump forms near the laser axis which is efficiently accelerated. The dense clump is about 1 - 2 laser wavelengths in diameter. The stabilization is

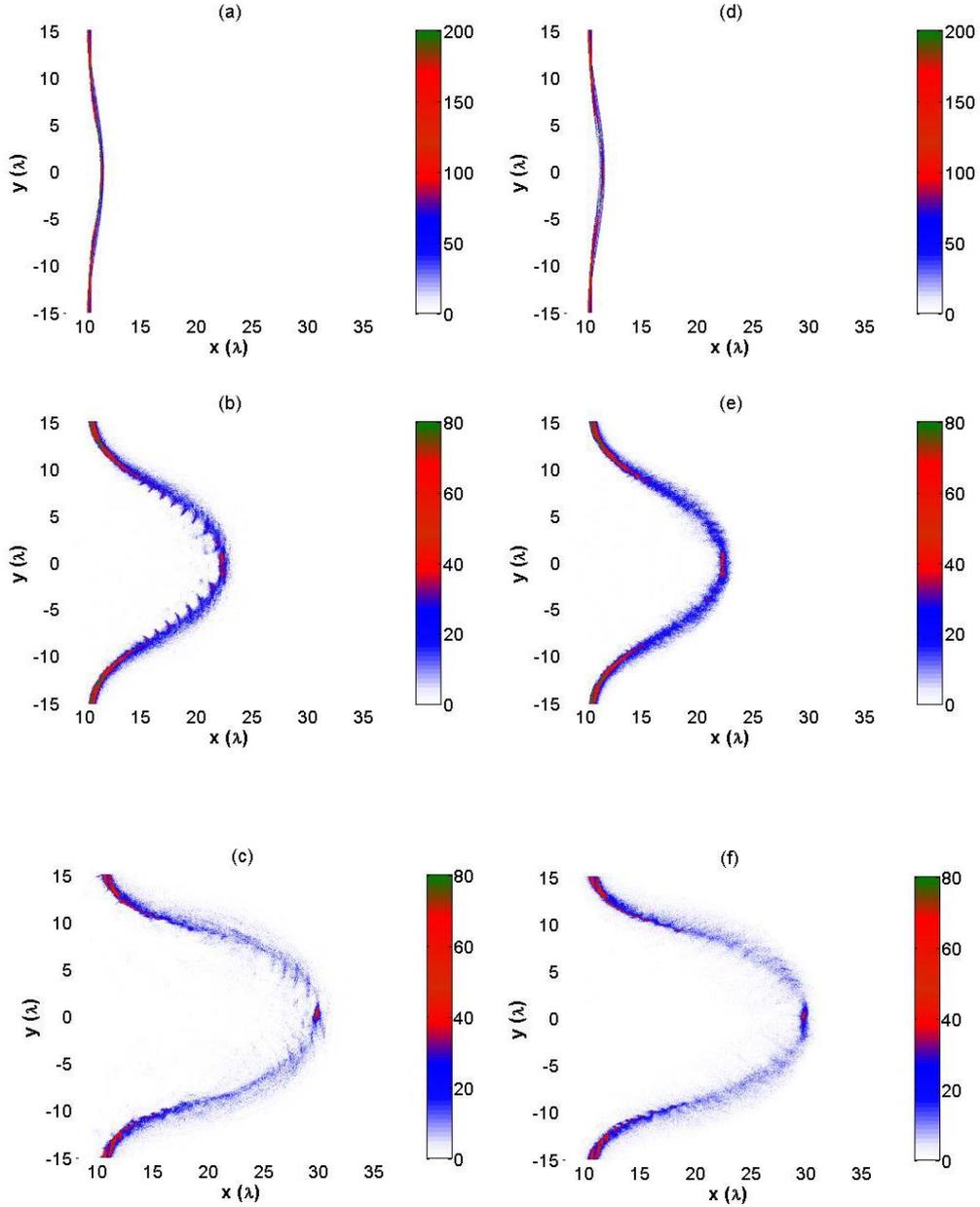

Fig. 1 (color online): Foil density evolution. Left: electrons, right: ions, at times (a,d) $t = 16$, (b,e) $t = 36$, (c,f) $t = 46$, in units of laser period. The laser pulse is incident from the left and hits the plasma at $t = 10$.

related to the driving laser pulse that has passed the dispersed foil in the transparent wing region and starts to encompass the opaque clump, keeping it together. Acceleration is then similar to that studied for so-called *reduced mass targets* [21], where small droplets or clusters are used as targets. In what is described below, the new configuration is self-organizing with small pieces of matter punched out of a plane foil. In this Letter, we exhibit this new regime in terms of two-dimensional particle-in-cell (2D-PIC) simulations.

In the simulations, we have taken a CP laser pulse with wavelength $\lambda = 1\mu$m and maximum normalized vector potential $a = eA/mc^2 = 50$, corresponding to an intensity of $I = 1.37 \times 10^{18}$ W/cm$^2 \cdot 2a^2/\lambda^2$. The pulse has a Gaussian radial profile with $20\lambda$ full width at half maximum and a trapezoidal shape longitudinally with $20\lambda$ flat top and $1\lambda$ ramps on both sides. It is normally incident from the left on a uniform, fully ionized hydrogen foil of thickness $D = 0.5\lambda$ and normalized density $N = n_e/n_{crit} = 80$, where the electron density $n_e$ is given in units of the critical density $n_{crit} = \pi m_e c^2/\lambda^2$ and $c$ is the velocity of light. Proton to electron mass ratio is $m_p/m_e = 1836$. The size of the simulation box is $60\lambda \times 40\lambda$ in (x,y) directions, respectively. We take 40 particles per cell per species and a cell size of $\lambda/80$. The flat plasma foil is located at $x = 10\lambda$ initially. Periodic boundary conditions are used for particle and fields in transverse direction, and fields are absorbed at the boundaries in longitudinal direction.

The temporal evolution of the foil is shown in Fig. 1, separately for electron and ion density. One observes that electrons and ions move closely together. At $t = 16$ (in unit of laser period), about 6 laser cycles after the pulse front has reached the plasma, the foil is slightly curved, following the transverse Gaussian profile of the laser pulse. At $t = 36$, a periodic structure having approximately 1 λ scale is seen, very prominently in the electron distribution, but also already imprinted in the ion distribution. Such surface rippling has been identified before in a number of numerical studies [6, 18, 19, 21-23] and has been described as a Rayleigh-Taylor-type instability (RTI) occurring in thin foils when driven by strong radiation pressure [24-26]. Here we depict it when the foil is already strongly deformed. At this time, the laser light is reflected from the indented walls and creates an intense standing-wave pattern at the bottom of the crater (see Fig. 2a).

We attribute the foil rippling to this λ-period seed pattern, at least in part. A second source is a fast current instability, setting in at early times when the foil is still plane. It has been described in [23]. Inspecting the longitudinal $j_x$ current at time $t = 36$ in Fig. 2c, a periodic structure of current cells can be recognized, also with λ period. It indicates a pattern of forward and backward currents typical for Weibel instability, which is known to grow fast on the time-scale of the inverse plasma frequency $\omega_p^{-1}$, which is shorter than the light period for solid density. These current patterns contribute to the unstable foil dynamics in the wing region.

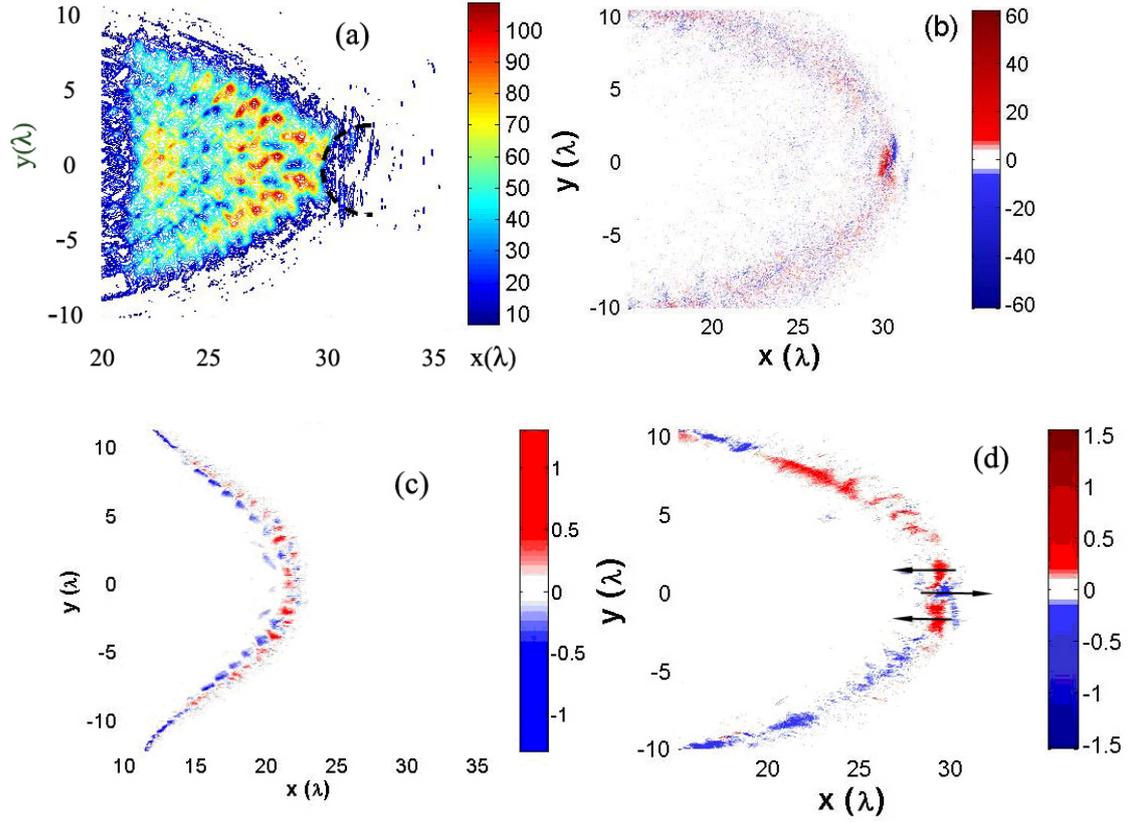

Fig.2 (color online) (a) laser field $\sqrt{E_y^2 + E_z^2}$ at $t = 46$, the dashed line marks the concave pulse front pushing the ion clump; (b) charge density distribution ($n_i - n_e$) at $t = 46$; (c) current density at $t = 36$ (normalized by $en_e c$); (d) current density at $t = 46$. The black arrows in the clump region indicate the direction of electron motion.

Another important aspect is that the foil is strongly accelerated as a whole, and RTI depending on ion motion will add to perforate and break the foil. This is clearly seen in Figs. 1c and 1f, both for electrons and ions. Using the RTI growth rates derived in [25], we find e-folding within 6 laser cycles for the wing region which is consistent with the present simulation. The actual dynamics are very complicated, combining RTI and Weibel instabilities in the phase of nonlinear evolution. This is not yet understood in detail and needs separate investigation. As a result, the foil becomes transparent in the wing region, where then light starts to pass the foil and to overtake the dense clump located near the laser axis (see Fig. 2a).

Let us now describe the evolution of the central clump in more detail. From Figs. 1c and 1f we see that the transverse extension of the clump is about $2\lambda$ at $t = 46$. It is driven by a laser field distribution having the form of a concave bowl which encompasses the clump and tends to keep it together. Plotting charge density $(n_i - n_e)$ in Fig. 2b, one recognizes how electrons (blue) are running ahead dragging ions (red).

The current cells, seen with $\lambda$-period at time $t = 36$ in Fig.2c, have almost been dissolved at time $t = 46$ in Fig.2d, except for the central cell on the laser axis which is stabilised by the local laser field. The black arrows in Fig.2d indicate the electron motion around the clump. Electrons move in laser direction on the axis and return on the side of the clump. This kind of current dipole may add to the stabilization of the clump by magnetic compression. Unfortunately, we did not succeed to show this B-field separately, because of the strong laser field superimposed.

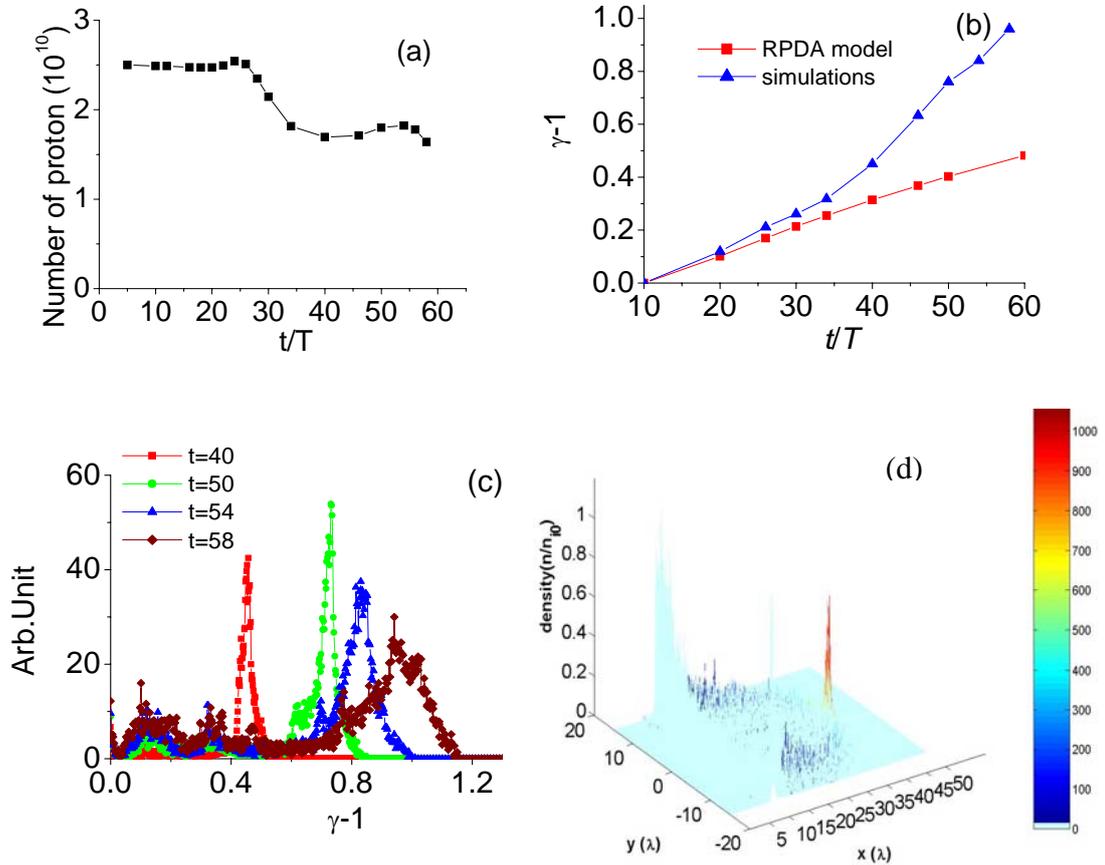

Fig.3 (color online) (a) Number of protons in the center of the foil ($r \leq \lambda/2$) versus time in units of laser cycles; (b) proton energy; (c) evolution of energy spectrum for beam ions located inside the central clump ($r \leq \lambda/2$); (d) energy distribution of protons at $t = 58$ (the colour bar gives ion energy in MeV).

Figure 3 highlights the central results concerning clump evolution. The total number of protons, comprised within a $\lambda/2$ distance from the laser axis and shown in Fig. 3a, drops after time $t = 26$ from an initial value of $2.5 \times 10^{10}$ due to transverse expansion, but this trend is interrupted at about $t = 35$, when the foil becomes transparent in the wing region and the new regime of quasi-stable acceleration sets in. In the present 2D-PIC simulation, about $1.7 \times 10^{10}$ protons (1 nano-Coulomb) are trapped in the central clump and are accelerated to an ion energy of approximately 1 GeV, as it is seen in Fig. 3b. An enhanced acceleration mode sets in at t=35 after clump formation, exceeding the predictions of the Radiation Pressure Driven Acceleration (RPDA) [27] by a factor 2. At the same time, the ion energy spectra start to exhibit sharp peaks, as it is seen in Fig. 3c. The perspective view in Fig. 3d then shows ion density in the $(x, y)$ plane with colour marking ion energy. One observes the high-density region of the unperturbed foil at the boundaries, the low-density plasma remains of the foil in the wing region (dark blue), and the accelerated ion clump sticking out as a conspicuous red spike in the centre. Results similar to those presented here have been obtained with other independent 2D-PIC codes [28]. Also 3D-PIC simulations reveal the existence of the new regime [29].

The present results depend on the high-contrast laser pulse (1 cycle rise time). More extended rise times tend to disperse the foil plasma not only in the wing region, but also to prevent ion trapping in the central dense clump. Although much progress has been made recently in generating laser pulses with ultra-high contrast [30], there may be other ways by modifications of the irradiation geometry to improve ion trapping. More systematic investigations based on 2D- and 3D-PIC simulation are now under way to map out the parameter space in which the new regime of central clump acceleration can be achieved.

In summary, we have identified a new regime of laser ion acceleration and have described the essential dynamics, self-organizing a mass-limited ion clump which is accelerated in a quasi-stable manner over many laser cycles without dispersion. This leads to sharply peaked proton spectra with energies of 1 GeV and more. These findings, obtained on the basis of multi-dimensional PIC simulations, go beyond previous results on phase stable acceleration published so far. An important point is that the nonlinear physics itself select the amount of accelerated ions (about 1 nano-Coulomb of protons in the present simulations), rather than relying on complicated target structures. This opens an option to use simple targets adequate for high-repetition rates by means of plane foils. This is attractive for applications.

**Acknowledgements**

One of the authors (XQY) would like to thank the Alexander von Humboldt Foundation for a scholarship. This work is partially supported by NSFC (10855001), by the Munich Centre for Advanced Photonics (MAP), and by the Association EURATOM – Max-Planck-Institute for Plasma Physics. ZMS is supported in part by the National Nature Science Foundation of China (Grants No. 10674175, 60621063). The authors acknowledge stimulating discussions with Prof. H. Ruhl and also with Dr. M. Chen and Prof. A. Pukhov.

- On leave Peking University

**References**

[1] G. A. Mourou, T. Tajima, and S. V. Bulanov, Rev. Mod. Phys. 78, 309 (2006).

[2] A. J. Mackinnon, Y. Sentoku, P. K. Patel, D. W. Price, S. P. Hatchett, M.H. Key, C. Andersen, R. A. Snavely, and R. R. Freeman, Phys. Rev. Lett. 88, 215006 (2002).

[3] M. Borghesi et al., Phys. Plasmas 9, 2214 (2002).

[4] S. V. Bulanov, T. Zh. Esirkepov, V. S. Khoroshkov, A. V. Kuznetsov, F. Pegoraro, Phys. Lett. A 299, 240 (2002).

[5] M. Roth et al., Phys. Rev. Lett. 86, 436 (2001); N. Naumova et al., Phys. Rev. Lett. 102, 025002 (2009).

[6] S. C. Wilks, W. L. Kruer, M. Tabak, and A. B. Langdon, Phys. Rev. Lett. 69, 1383 (1992).

[7] E. L. Clark et al., Phys. Rev. Lett. 85, 1654 (2000); A. Zhidkov et al., Phys. Rev. E 60, 3273 (1999).

[8] J. Denavit, Phys. Rev. Lett. 69, 3052 (1992).

[9] A. Zhidkov, M. Uesaka, A. Sasaki, H. Daido, Phys. Rev. Lett. 89, 215002 (2002); L. O. Silva, M.Marti, J.R.Davies, R.A.Fonseca, C.Ren, F.Tsung, W.B. Mori, Phys. Rev. Lett. 92, 015002 (2004); A. Maksimchuk, S. Gu, K. Flippo, D. Umstadter, V.Y. Bychenkov, Phys. Rev. Lett. 84, 4108 (2000).

[10] P. Mora, Phys. Rev. Lett. 90, 185002 (2003); T. Esirkepov, M.Yamagiwa, and T.Tajima Phys. Rev. Lett. 96, 105001 (2006).

[11] Y. T. Li, Z. M. Sheng et al, Phy. Rev. E 72, 066404 (2005).

[12] V. Malka et al., Med. Phys. 31, 1587 (2004).

[13] H. Schwoerer et al., Nature 439, 445 (2006).

[14] T. Toncian et al., Science 312, 410 (2006).

[15] M.Hegelich et al, Nature 439, 441 (2006).

[16] A. Macchi, F. Cattani, T.V. Liseykina, F. Cornolti, Phys. Rev. Lett. 94, 165003 (2005).

[17] S.G. Rykovanov, J. Schreiber, J. Meyer-ter-Vehn, C Bellei, A. Henig, H.C. Wu and M. Geissler, New J. Phys. 10, 113005 (2008) ; X. Zhang, et al., Phys. Plasmas 14, 123108 (2007); C.S. Liu, V.K. Tripathi and X. Shao, Frontiers in Modern Plasma Physics 246-254 (2008).

[18] O. Klimo, J. Psikal, and J. Limpouch, and V. T. Tikhonchuk, Phys. Rev. Special Topics - Accelerators and Beams 11, 031301 (2008).

[19] A. P. L. Robinson, M. Zepf, S. Kar, R.G. Evans, and C. Bellei, New J. Phys. 10, 013021 (2008).

[20] X.Yan et al., Phys. Rev. Lett. 100, 135003 (2008).

[21] A. Henig, D. Kiefer, M. Geissler, S.G. Rykovanov, R. Ramis, R. Hörlein, J. Osterhoff, Zs. Major, L. Veisz, S. Karsch, F. Krausz, D. Habs,and J. Schreiber, Phys. Rev. Lett. 102, 095002 (2009).

[22] A. Macchi, F. Cornolti, F. Pegoraro, T.V. Liseikina, H. Ruhl, and V. A. Vshivkov, Phys. Rev. Lett. 87, 205004 (2001).

[23] M. Chen, A. Pukhov, Z.M. Sheng, X.Q.Yan, Physics of Plasma, 15, 113103 (2008).

[24] E. G. Gamaly, Phys. Rev. E 48, 2924 (1993).

[25] F. Pegoraro and S.V. Bulanov, Phys. Rev. Lett. 99, 065002 (2007).

[26] Y. Yin, W. Yu, M. Y. Yu, A. Lei, X.Yang, H. Xu and V.K. Senecha, Phys. Plasmas **15**, 093106 (2008).

[27] T. Esirkepov, M. Borghesi, S. V. Bulanov, G. Mourou, and T. Tajima, Phys. Rev. Lett. 92, 175003 (2004).

[28] H.C. Wu, Z.M. Sheng, J. Zhang, Phys. Rev. E 77, 046405 (2008).

[29] The formation of a central clump and proton acceleration to GeV energies is also observed in 3D PIC simulations, when using similar parameters as in the present paper. M. Chen, private communication.

[30] R. Hoerlein, et al, New Journal of Physics 10, 083002 (2008).